\documentclass[a4paper]{jpconf}
\usepackage{graphicx}
\usepackage{hyperref}
\hypersetup{
    colorlinks,
    citecolor=black,
    filecolor=black,
    linkcolor=black,
    urlcolor=black
}
\newcommand{\STQ}{{Space Time Quest}}
\newcommand{\BHP}{{Black Hole Pong}}
\newcommand{\GW}{{GW}}
\newcommand{\gwoptics}{\href{http://www.gwoptics.org}{gwoptics.org}}

\begin{document}
\title{Computer-games for gravitational wave science outreach: {\it Black Hole Pong} and {\it Space Time Quest}}
\author{
L.\,Carbone, 
C.\,Bond, 
D.\,Brown, 
F.\,Br\"uckner, 
K.\,Grover, 
D.\,Lodhia, 
C.M.F.\,Mingarelli, 
P.\,Fulda, 
R.J.E.\,Smith, 
R.\,Unwin, 
A.\,Vecchio, 
M.\,Wang, 
L.\,Whalley
and A.\,Freise}

\address{School of Physics and Astronomy, University of Birmingham\\ Edgbaston - Birmingham, B15 2TT, United Kingdom}

\ead{lc@star.sr.bham.ac.uk}

\begin{abstract}
We have established a program aimed at developing computer 
applications and web applets to be used for educational purposes as well 
as gravitational wave outreach activities. These applications and applets 
teach gravitational wave physics and technology.
The computer programs are generated in collaboration with 
undergraduates and summer students as part of our teaching activities, and are freely distributed on a dedicated website.
As part of this program, we have developed two computer-games related to gravitational wave science: 
`Black Hole Pong' and `Space Time Quest'. 
In this article we present an overview of our computer related outreach 
activities and discuss 
the games and their educational aspects, 
and report on some positive feedback received. 
 \\ {\bf To be published on:} \href{http://iopscience.iop.org/1742-6596}{\it Journal of Physics:\,Conference Series}, 
 {\it Proceedings of the 9$^{th}$ Amaldi Conference on Gravitational Waves, Cardiff 2011}
\end{abstract}

\section{Introduction}
The next generation of gravitational wave (\GW) detectors - most
notably Advanced LIGO \cite{aligo}, GEO-HF \cite{geohf} and Advanced VIRGO \cite{avirgo} -
are expected to achieve the first direct detection of \GW s 
 during 2015, when Advanced LIGO begins operation or soon thereafter, 
giving life to the new field of the
\GW~astronomy. 
The impact on cosmology, fundamental physics, and of course on
conventional astronomy itself will be enormous; the interest in
\GW~observation is therefore 
growing rapidly in the scientific community. But while the public
interest in cosmology and relativity is high, public knowledge of the
principles behind \GW~detection is still rather limited, and even the
existence of the large \GW~observatories is little known by the more
general public, compared to larger experimental facilities such as,
for instance, the Large Hadron Collider \cite{LHC}. There is a clear
need to better inform and inspire the general public and prospective
students in \GW~astronomy and related sciences.
Within the LIGO Scientific Collaboration (LSC) 
the `Education and Public Outreach' (EPO) group aims to combine ideas and
approaches across the collaboration to successfully communicate the
vision and benefits of GW observation  throughout the world.

To contribute to these international efforts in the promotion of GW
astronomy, at the University of Birmingham we have established a
unique program aimed to the development of small educational
computer applications that can be used to illustrate the basics of
\GW~science and \GW~detector technology in a playful but informative
way. The aim is to present GW science to younger generations within
one of the environments they are more familiar with,  i.e. computer-games, 
and to exploit the communication channels that the {\it new
  technologies} offer to possibly get an even larger international
audience in contact with \GW~science. 

This activity led us to the successful development of a number of
interactive computer applets describing a variety of concepts
connected to \GW~science and, eventually, to the realisation of
two full-scale computer-games based on the subjects of gravity and \GW
s. In this article we overview our computer related outreach activity
and present and discuss our \GW~related games, \BHP~and \STQ.

\section{{`Processing'} programs for science outreach}
The idea of developing small computer applications for educational 
purposes builds upon the need to introduce new students to the world of 
computer programming and modelling of physical systems in a manner 
which is enjoyable. This will not only help them learn successfully, but 
also engages them with GW research.
During the initial `induction' 
phase, undergraduate or summer-students involved in research projects
with our group are encouraged to generate a small computer program on
a \GW~subject of their interest, which is then developed, as a
conventional student-project, by the student with the supervision of
more senior members of the group. 

The small computer programs, called `sketches', are developed using
the open-source programming environment Processing \cite{processing:book,processing:web}, 
originally developed at the MIT Media Lab in 2001 as a software prototyping environment and to teach
fundamentals of computer programming within a graphical context.
Processing has eventually reached a wide audience
and is now largely used in many professional communities, such
as designers, artists, and architects to create graphical
applications, animations, interactive tools and for visual arts in
general. 
Processing offers an intuitive approach to programming for
the beginner or an efficient sketchbook for rapid prototyping by
experienced programmers. Thus Processing allows students with very different
computing-backgrounds to collaborate and to successfully produce graphically
impressive sketches in a relatively short period of time. 

The successful sketches are published online on our outreach website
\gwoptics~\cite{gwoptics}, on individual webpages where the
student-author can provide instructions and a short description of the
physics illustrated in the sketch. The program is either embedded
within the HTML code to run as an applet in the webpage itself or,
where more appropriate, distributed for download as an application to
install and run on the computer of the interested person.

The collection of sketches developed so far covers a wide spectrum of
subjects related to \GW~science and technologies. The programs range
from illustrating the most fundamental properties of \GW s, for
example the deformation of space-time produced by a propagating \GW~or
the characteristic `sound' of the \GW~signal of colliding black holes,
to the illustration of the vital technologies and phenomena used in GW
detectors such as lasers, vibration isolation systems or interference
of light. Different combinations of these sketches have been
successfully used as interactive tools during seminars about GW
detection and during more general lectures in schools and
universities, and the sketches webpages are regularly consulted online
by people interested in learning about the specific subject.

\begin{figure}[th]
\begin{center}
\includegraphics[width=7.8cm]{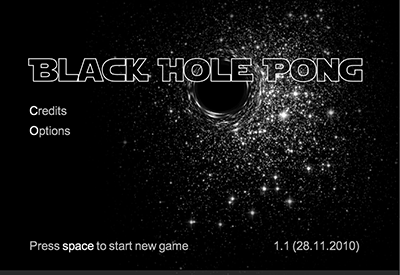}
\includegraphics[width=7.8cm]{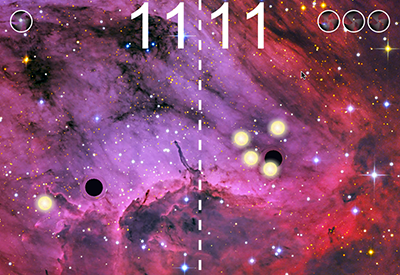}
\caption{\label{fig:BHPscreenshots} Screen-shots of the \BHP~game. The
  image on the left shows the start-up screen of the game and the
  right snapshot has  been 
  taken mid-game, showing one of the
  astronomical background images, the two `black holes' controlled
  by the players and several bright disks as the stars currently
  in play.}
\end{center}
\end{figure}

\section{The gravitational wave games}
The positive feedback we received about the different interactive sketches has 
encouraged us to realise two properly defined computer-games 
based on subjects related to gravity and GW science. 
The motivation behind the development of the two particular games was
as follows:
in one case, the aim was to produce an intuitive and graphically
attractive computer-game that could engage and entertain children and
teenagers during science exhibitions;  in the other, 
the goal was
vice-versa to develop an interactive element that could function as an
engaging and playful supplement for illustrating and explaining the
secrets of \GW~detectors and their technology to a more educated
audience, such as high-school students onwards. The two games, \BHP~and
\STQ, are presented and discussed in this section.

\subsection{Black Hole Pong}
\BHP~(BHP) is a new arcade-style game with a reference to 
one of the very first computer-games, {\it Pong}~\cite{Pong}.
Pong involved two players, each one controlling a paddle
which they would move vertically up and down in order to bounce a 
ball back towards their opponent: each
time the ball touched the opponent's far edge of the screen, the player
would score a point. In BHP the idea of the split-screen has been
kept. However each player controls a black hole which can move
horizontally as well as vertically, and the
objective is to make use of the gravitational potential of the black
hole to fling free roaming stars towards the other player.
\begin{figure}[b]
\begin{center}
\includegraphics[height=4.1cm]{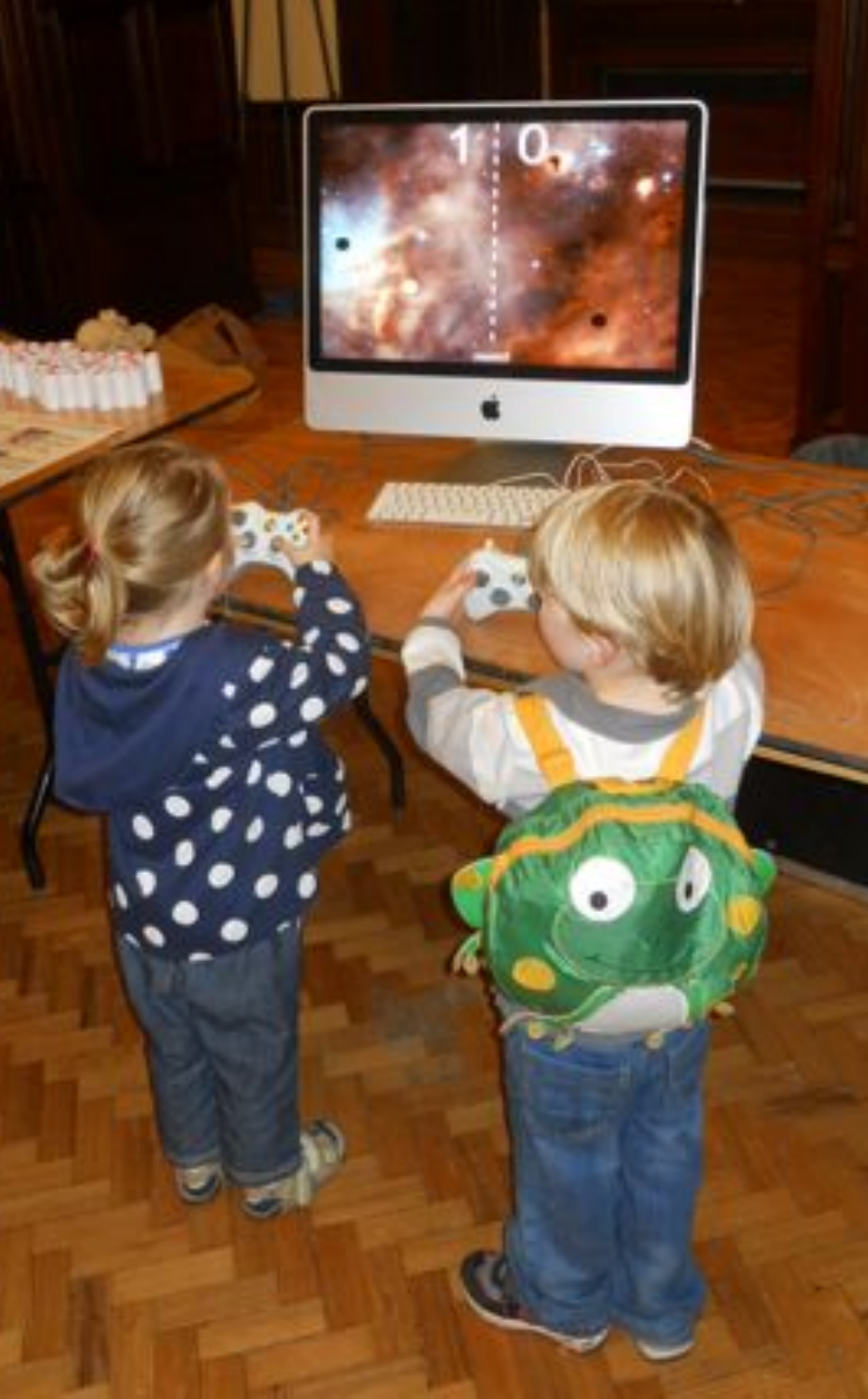}
\includegraphics[height=4.1cm]{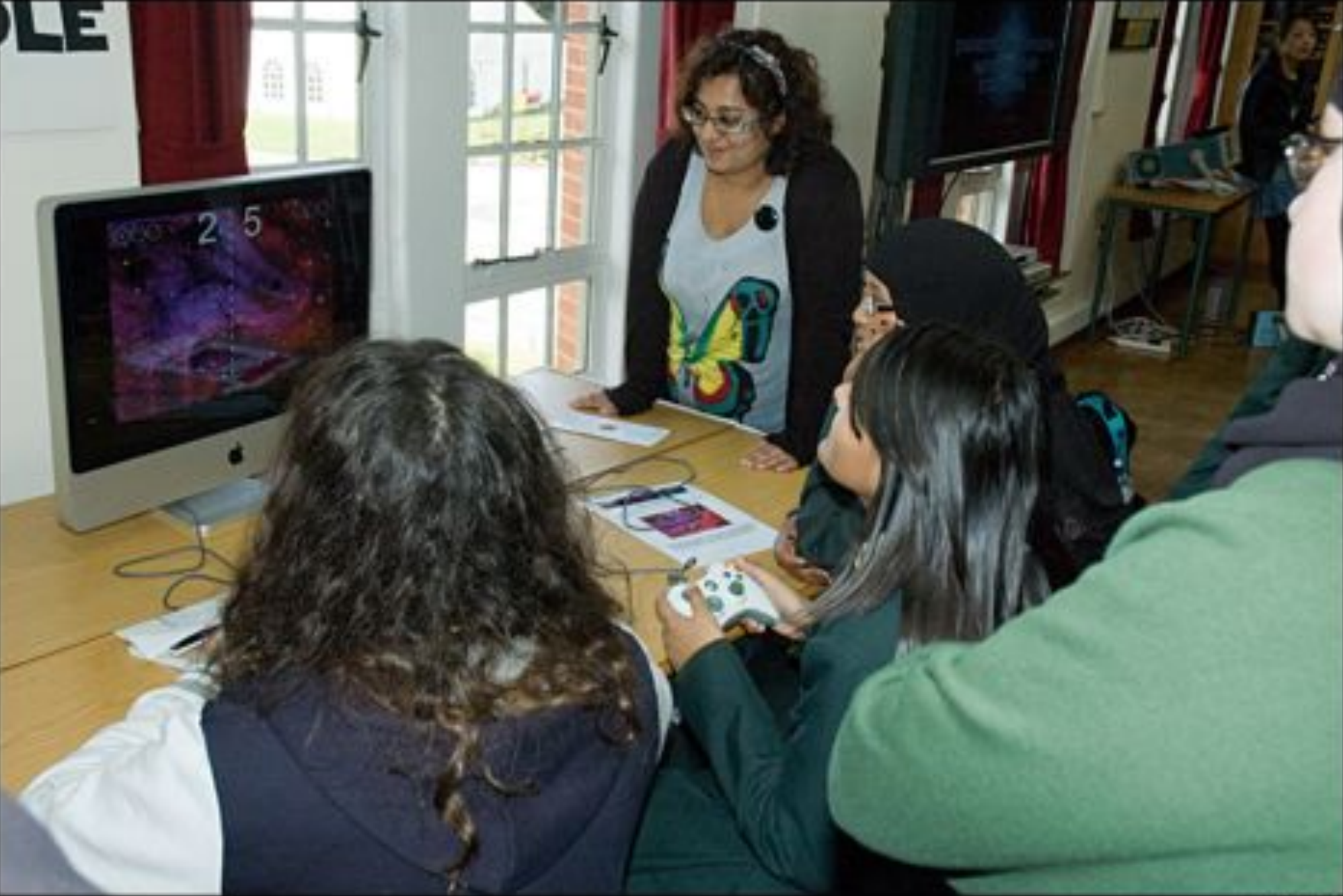}
\includegraphics[height=4.1cm]{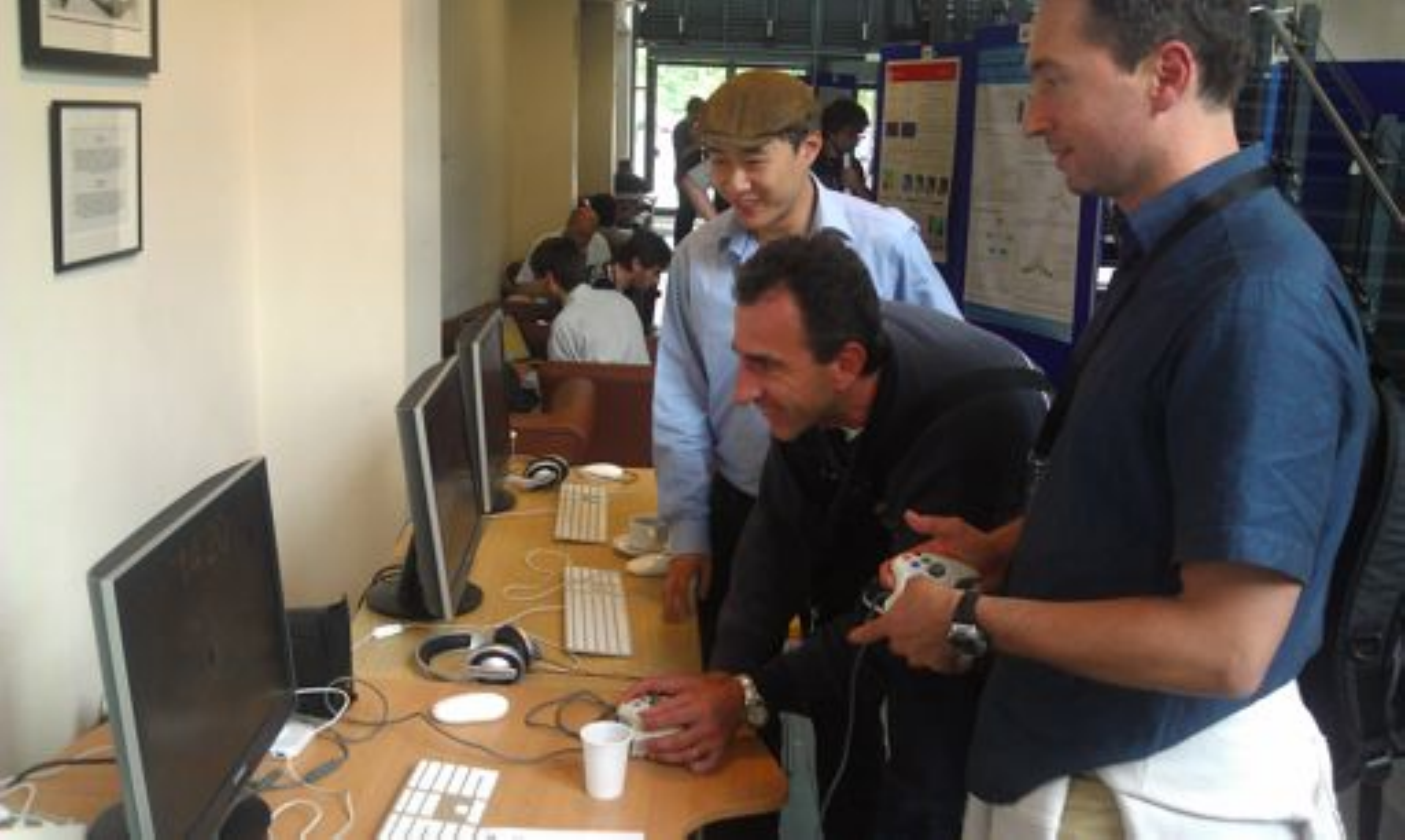}
\caption{\label{fig:BHPexhibition} Children, students and adults enjoying the BHP~game 
at the University of Birmingham Community Day 2011 (left), the British
Science Festival 2010 (center) and the 9$^{\rm th}$ Amaldi Conference
(right).}
\end{center}
\end{figure}

BHP~has been designed  as a simple, fun game for people of
all ages. At the same time, BHP features several educational elements
that make it a fascinating tool for teaching, learning and discovering
new physical concepts. For example, by learning how to manoeuvre the
incoming stars only using the black hole's gravitational potential,
the player develops an intuitive understanding of concepts such as
gravitational attraction, orbital
mechanics and gravitational slingshot effect. 
The background graphics of the game is a slideshow containing images of 
the night sky as well as astronomical objects taken by both amateur 
astronomers and large ground/space based telescopes.
Furthermore, several astrophysical phenomena are graphically featured
in the game, such as `worm holes', `star capture'
and `gravitational lensing', adding to the overall simple but
attractive graphics, see Fig.~\ref{fig:BHPscreenshots}.
When used as part of an exhibition, science fair or similar activity we have 
supported the arcade-style attraction by state-of-the art hardware:
the game is designed to be controlled with the well-known Microsoft
Xbox controllers and whenever possible we use large Apple iMac
computers to run the game, as shown in Fig.~\ref{fig:BHPexhibition}.

\subsection{Space Time Quest}
\begin{figure}[t]
\includegraphics[height=5.3cm]{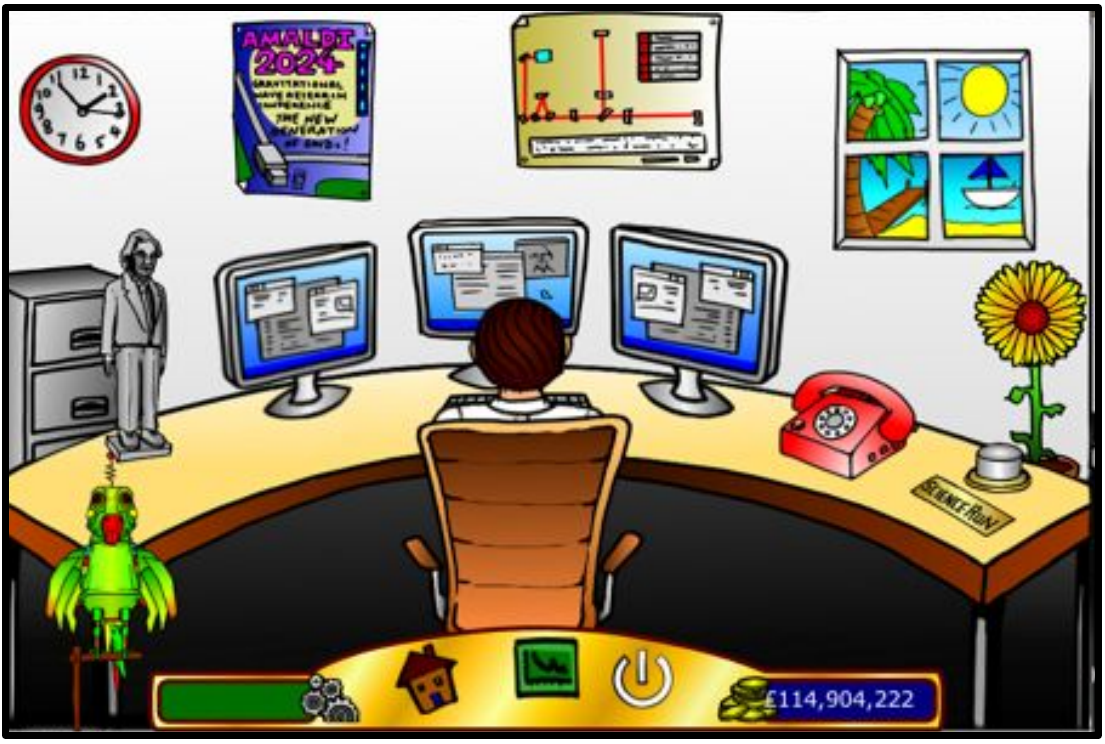}
\includegraphics[height=5.3cm]{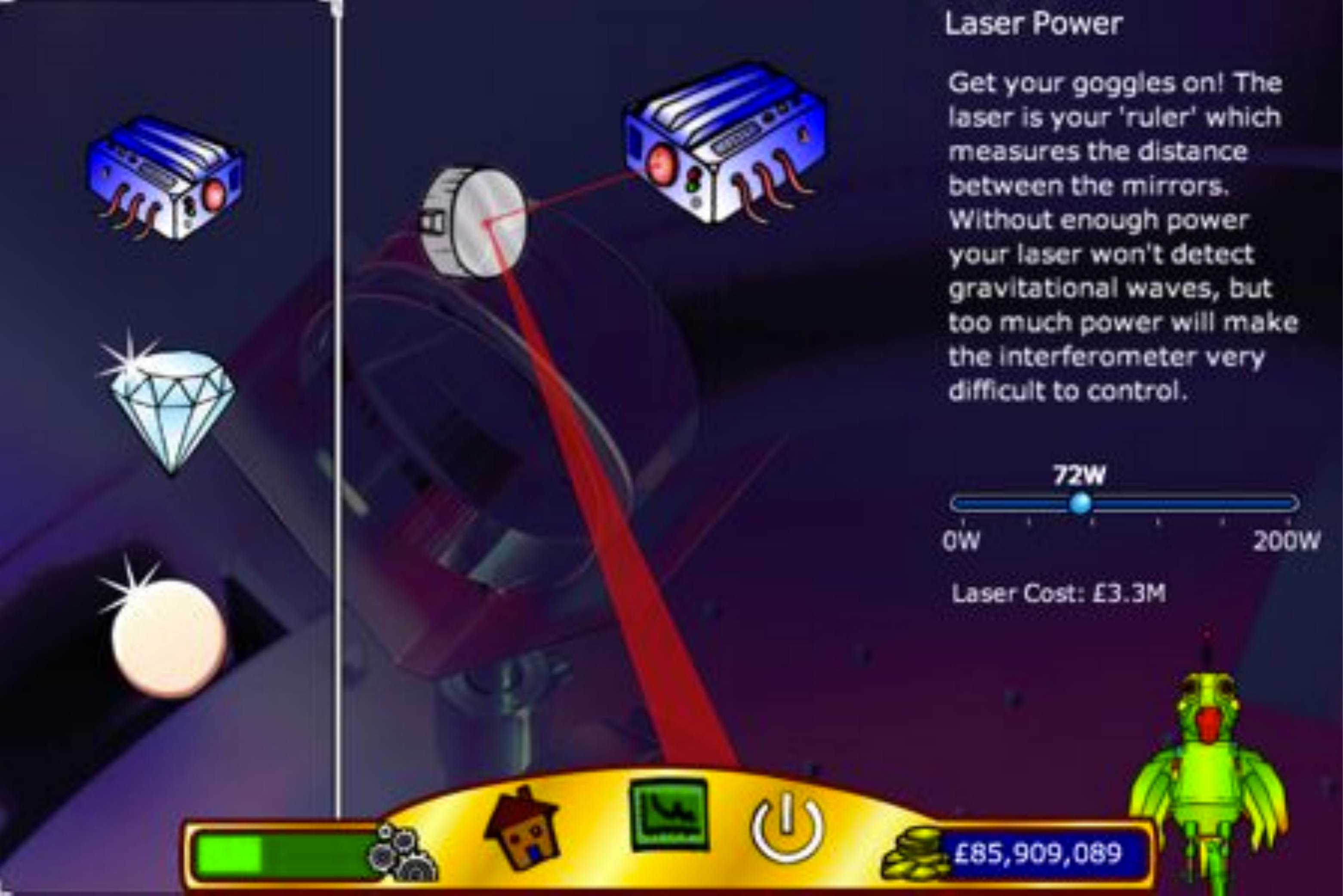}
\caption{\label{fig:STQ:screens} Screenshots of STQ. On the left, the `PI's office' 
is the game's hub from which the player can access all the detector's subsystems. On the right an image of the `Optics' subsystem screen.}
\end{figure}

\begin{figure}[t]
\includegraphics[height=5.3cm]{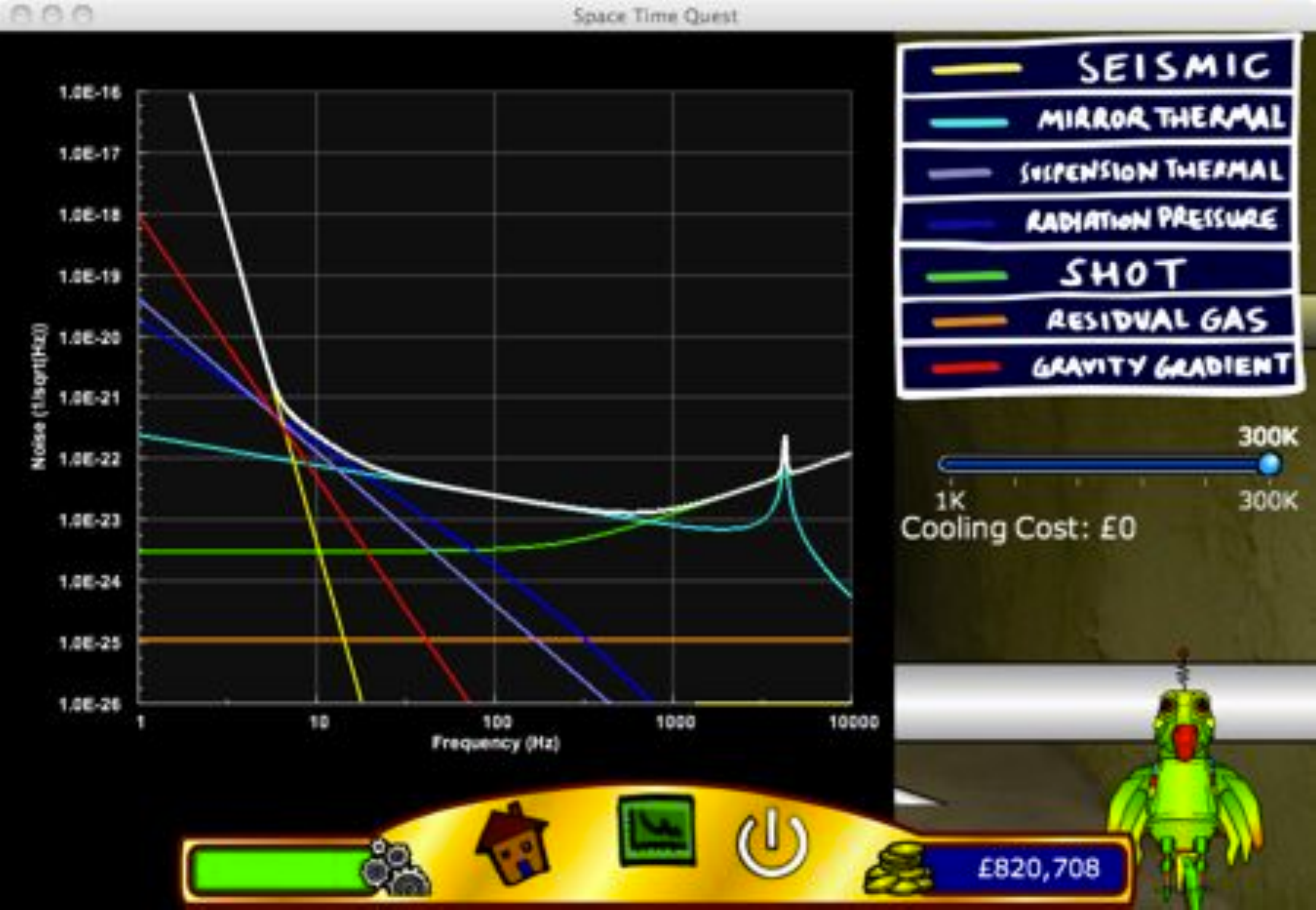}
\includegraphics[height=5.3cm]{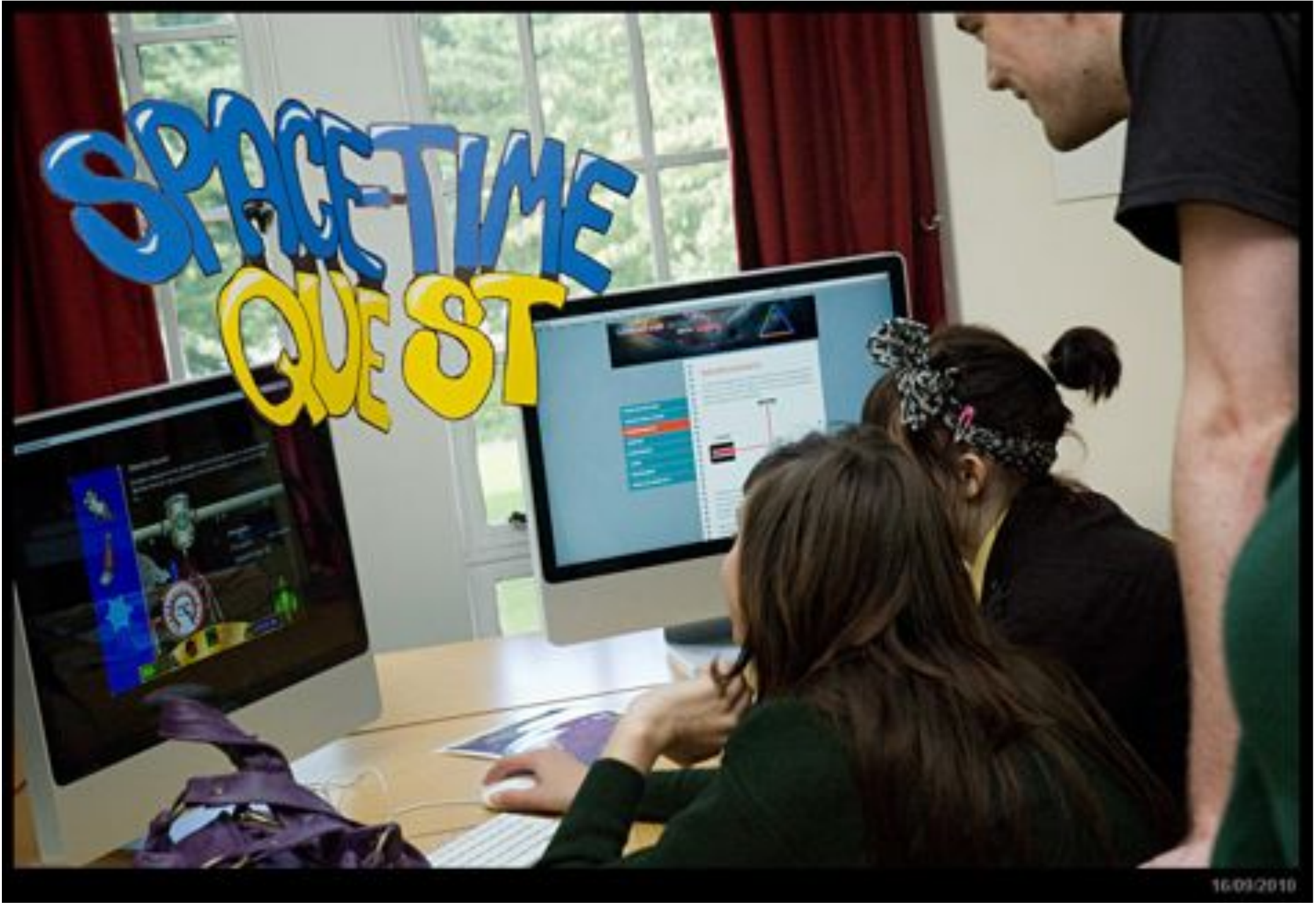}
\caption{\label{fig:STQ:noise} Left: a screen-shot of the `sensitivity curve' in the STQ game. 
Right: school students playing 
STQ with a demonstrator during an exhibition at University of Birmingham.}
\end{figure}

\STQ~(STQ), here shown with screenshots in Fig.~\ref{fig:STQ:screens} and Fig.~\ref{fig:STQ:noise}, is a manager-simulation type game: 
the player is the `Principal Investigator' (PI) of a 
future ground-based
\GW~interferometer who has the goal to 
design 
the most sensitive GW detector.
The PI is assigned a limited budget that he has to wisely distribute between 
the different detector's subsystems 
to tweak the instrument parameters and to achieve 
the best sensitivity. Once the player is satisfied with their design, 
they can operate the detector in `Science-Run' mode and see how well it performs:
the final score is determined by how far in the universe the detector itself can explore, 
based on the achieved sensitivity curve,
and it is recorded in the web-based STQ `hall-of-fame'.

STQ is complemented by the \href{http://www.gwoptics.org/ebook}{`Gravitational Waves E-Book'} \cite{ebook}, 
effectively a collection of webpages with short introductions to a number of topics relevant to GW science. 
The E-Book offers a description of the main instrument subsystems that comprise the detector and 
illustrates 
how each noise source relates to the
different subsystem parameters that the player chooses. 
The E-Book text is purposely aimed to
the general person who has interested in GW science,
and as such it is written with a simple style 
to make it accessible 
to a broad public of all 
ages and backgrounds.
The E-Book is also independently available online as more general reading material on GW science, 
and is now also offered in multiple languages.

STQ contains many educational merits for the public. 
First of all, the game showcases the physics behind a real 
GW detector and it presents all the main subsystems that comprise it. 
Furthermore, it illustrates all the most important noise sources 
which can limit the detector's sensitivity. 
By reading the E-Book and by looking at the changes in the sensitivity curve, 
the player can see 
how each subsystem is affected by the different noise sources, 
and discover some of the ways in which physicists 
try to reduce the noise sources in the detector. 
The game also presents some of the typical challenges that physicists face when designing 
a real physics experiment,
such 
as 
making trade-offs between performances of different 
interlinked subsystem parameters and managing the available resources wisely. 
Finally, STQ features images of astronomical objects in the background, 
similar to BHP. Furthermore 
the STQ graphics are based on photographs of components from real GW detectors complemented
with realistic cartoons of the detector parts, offering a realistic picture of how a GW interferometer looks like to the user.

STQ~is mainly targeted for science teachers, A-level, Higher and
Advanced Higher science students and is mainly suited for use in
science fairs and exhibitions and initially played with the help of
demonstrators. However, STQ~has also proved to be an entertaining
tool to teach the basics of \GW~science also to beginners in
\GW~research-projects and PhD schools, or to attract prospective
research students towards the \GW~field.

\section{Use in exhibitions and distribution}
\begin{figure}[b]
\includegraphics[width=16cm]{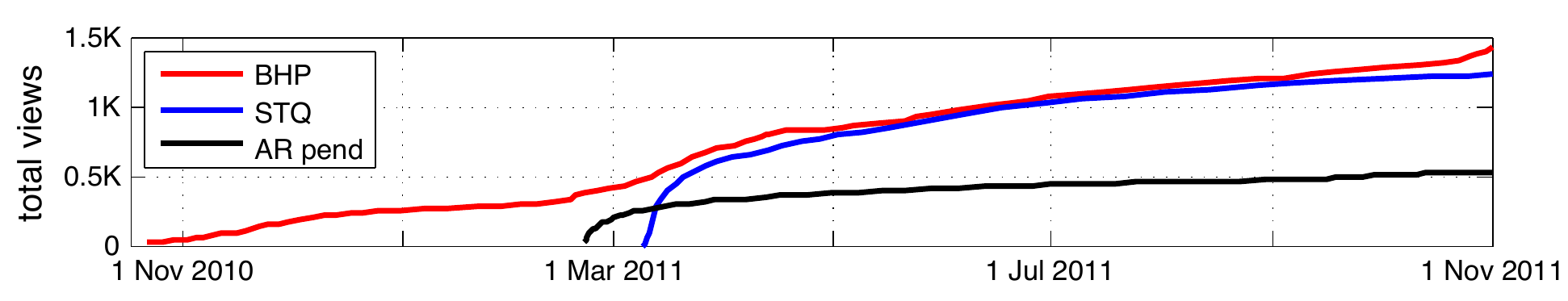}
\caption{\label{fig:youtube} Total number of unique views of the online video tutorials for the Black Hole Pong and Space Time Quest games and for one of our processing sketches, the `Augmented Reality Pendulum'. Data from \href{http://www.youtube.com/gwoptics}{youtube.com}.}
\end{figure}

Early prototypes of the games have been used for the first time 
during the exhibition about \GW~science `Looking
for Black Holes with Lasers', organised by the Birmingham GW Group
within the `British Science Festival', held in Birmingham in September
2010 
\cite{BSF:main}. 
The very positive feedback collected with the games
during this first exhibition gained us the attention of our university
and of local schools and associations. This led to our displays being
routinely used in University Open/Admission days and in our
university's outreach events, e.g. the University Community Day 2011
\cite{comday}, and gained members of our group invitations to visit
schools and give public seminars, where the games were used 
to complement the seminar. 
At the same time, positive feedback has been received from school 
teachers concerning our other Processing sketches which have been used 
as support material in physics lectures and during science activities.

\begin{figure}[t]
\includegraphics[width=16cm]{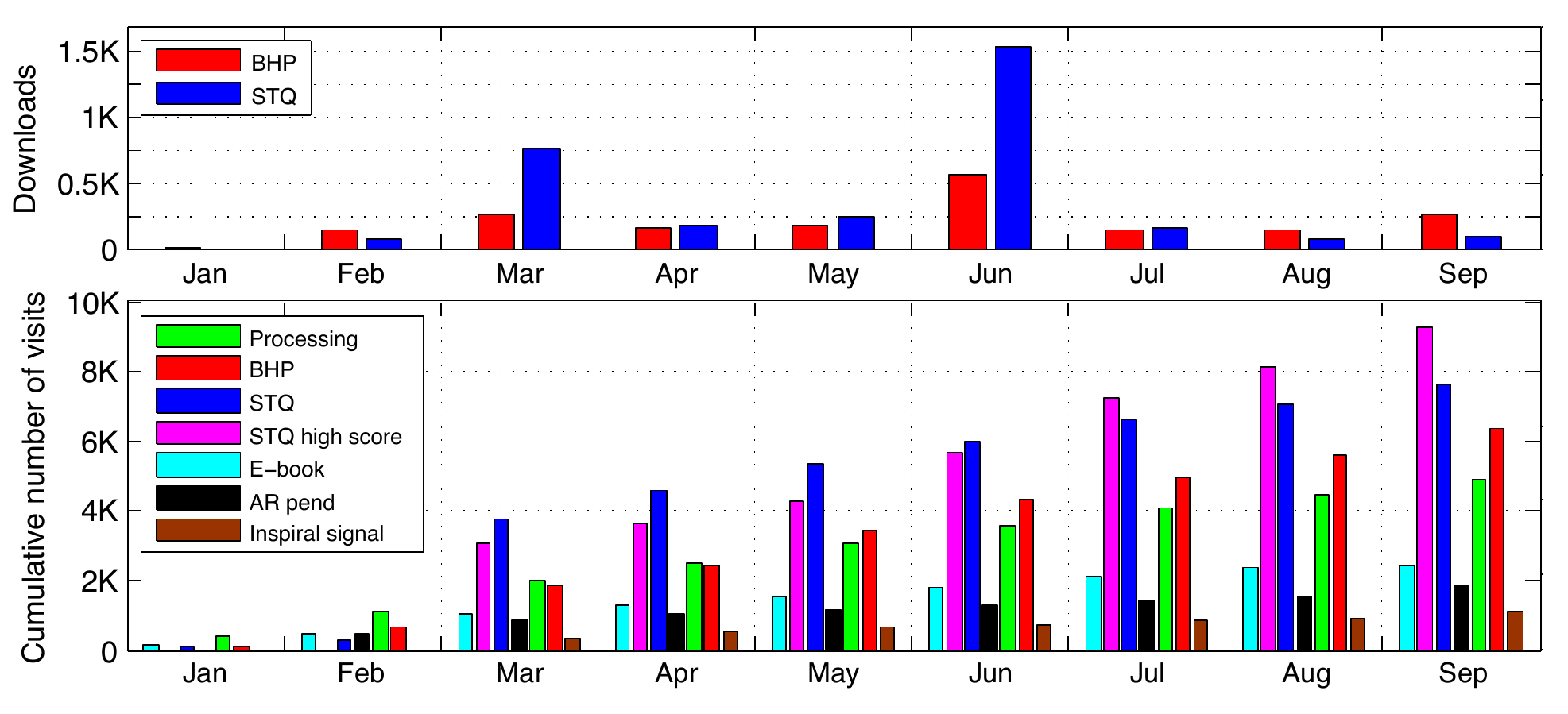}
\caption{\label{fig:gwoptics} Top panel: number of download of the Black Hole Pong and Space Time Quest games over the year 2011. Bottom panel, cumulative number of visits 
during 2011 to some of the \gwoptics~outreach pages, such as the main page collecting all Processing programs, the games BHP and STQ, the STQ high-score `hall of fame', the E-Book and the pages of two other Processing sketches, the `Augmented Reality Pendulum' and the `Inspiral signal'. }
\end{figure}

Since their official release, BHP and STQ are freely distributed on
our website \gwoptics~ and on the outreach pages of the \href{http://www.ligo.org/}{ligo.org} website \cite{ligo.org}, which hosts links and multimedia material of interest for the EPO group. The launch of the two games was also announced online via
social-media networks and with online videos, with the main aim 
of raising both the profile of the games and its 
visibility.
Indicative figures of merit
on how positive this campaign has been
can be inferred from
the total number of 
download 
of the two games, the entries in the
high score `hall of fame', the unique views of the online
video-tutorials, and more generally from the number of visits to 
the webpages presenting our online material. 
Examples of such data are presented in Fig.\,\ref{fig:youtube} and
Fig.\,\ref{fig:gwoptics}. 
The response is so far very positive and
seem to indicate a slow but constant growth of new contacts and an
increasing interest in the products themselves. In particular,
sensible increments in the number of contacts can be successfully
correlated with our contribution at science events and exhibitions,
with release of new outreach material and with our communication
campaign via online social networks.

STQ and BHP are also accompanied by short questionnaires, handed-out at
exhibitions and during visits to schools as well as online, 
which are distributed to collect anonymous feedback amongst the users,
targeting in particular teacher's and student's categories. 
The aim of the questionnaires is primarily to evaluate 
the success of the games among the users and in particular about the
effectiveness of their educational aspects. 
As a further step in the future, 
the goal is to develop a proper analysis of the feedback provided in the questionnaires 
that will allow us to better link the games to specific elements of education, such as formal education, 
and to improve their integration within the syllabus.

\section{Conclusion and future activities}
Our program aimed at the development of small computer applications 
for educational purposes. This has been successful, with the realisation of 
several interactive sketches and two full scale games
related to GW science and technology. Thanks to our participation 
at 
popular science events, and to our online presence and communication
campaign, the sketches and the games are now becoming popular within
schools and science associations and, as shown by the 
response gathered from the online audience, the prospective feedback looks
promising for the future. In particular,  
BHP and STQ  
have allowed us to significantly increase the visibility of
our online presence and as such also of 
GW 
science
within the general public, as well as within the local scientific community.

Encouraged by these positive results, we will continue our
computer related outreach activity in the next years, and we plan to
realise new Processing sketches for outreach in the near future. 
In parallel, BHP and STQ will be treated as running projects. We will take 
advantage of the feedback and advice from teachers, students and other 
users to make further modifications and improvements to both games.
Furthermore, we
will continue presenting BHP, STQ and our other interactive sketches
during visits to local schools and in popular science events. We hope
to increase and improve our online communication campaign
on GW subjects to make GW detectors more and more popular among the
general public and to gain GW science the largest audience possible. 

\ack 
We are very grateful to the Processing community for all the online examples, code libraries and online forums
that helped us in the development of our sketches and games and without whom most of this work would not have been possible. 
We thank the astronomers who allowed us to use their impressive photographs of the night sky as background images for the BHP and STQ games (see the credits page in the games for the details) and we are thankful to the GEO600 collaboration, for providing images of components of the detector used in the STQ game and for extensive beta-testing of the initial game prototype. This document has been assigned the LIGO Laboratory document number LIGO-P1100145.

\end{document}